\documentclass[aps,preprintnumbers,prl,twocolumn,groupedaddress,floatfix]{revtex4}

\usepackage{graphicx,amssymb}

\def\be{\begin{equation}}
\def\ee{\end{equation}}
\def\bea{\begin{eqnarray}}
\def\eea{\end{eqnarray}}

\def\bbuildrel#1_#2^#3{\mathrel{\mathop{\kern 0pt#1}\limits_{#2}^{#3}}}
\def\slash#1{\setbox0=\hbox{$#1$}#1\hskip-\wd0\dimen0=5pt\advance
       \dimen0 by-\ht0\advance\dimen0 by\dp0\lower0.5\dimen0\hbox
         to\wd0{\hss\sl/\/\hss}}

\newcommand{\gae}{\lower 2pt \hbox{$\, \buildrel {\scriptstyle >}\over {\scriptstyle
\sim}\,$}}
\newcommand{\lae}{\lower 2pt \hbox{$\, \buildrel {\scriptstyle <}\over {\scriptstyle
\sim}\,$}}

\begin{document}

\title{Generation of Noise Time Series with arbitrary Power Spectrum}

\author{M.~Carrettoni$^{1,2}$ and O.~Cremonesi$^{2}$}
\affiliation{$^{1}$Universit\'a degli studi di Milano Bicocca, Piazza della scienza,3, I-20126 Milano \\
$^{2}$INFN - Sez. Milano Bicocca, Piazza della scienza,3, I-20126 Milano}

\begin{abstract}
Noise simulation is a very powerful tool in signal analysis helping to foresee the system performance in real experimental situations. Time series generation is however a hard challenge when a robust model of the noise sources is missing. We present here a simple computational technique which allows the generation of noise samples of fixed length, given a desired power spectrum. A few applications of the method are also discussed. 

\end{abstract}

%\pacs{jinks}
% Fermilab Preprint FERMILAB-PUB-08-073-T

\maketitle
\section{Introduction} 
A standard approach in noise time series simulation is based on Carson's theorem \cite{carson31}. In fact, the superposition of randomly delayed pulses of a definite shape $f(t)$ with arbitrary coefficients a$_k$ gives rise to a pulse train 
\begin{equation}
n(t)=\sum_{k} a_k f(t-t_k)
\label{series}
\end{equation}
whose power spectrum $N(\omega)$ can be expressed in terms of the $f(t)$ power spectrum $P_f(\omega)$ according to
\begin{equation}
N(\omega)=\alpha P_f(\omega) = \alpha {|\widehat F(\omega)|^2}
\label{carson_law}
\end{equation}
where $\alpha$ is a proper normalizing constant and $\widehat F(\omega)$ is the Fourier transform of f(t). The delays $t_{k}$'s are distributed according to Poisson statistics and $f(t)$ is usually required to have a zero mean value so that $n(t)$ averages to zero. It can be shown that this result is independent of the density function used for the random amplitudes $a_k$, if the  variance of such distribution is fixed by the choice of the normalizing constant $\alpha$.\\

The basic idea of the method consists then in searching a proper expression for $f(t)$ so that the generated time series $n(t)$ have the desired  power spectrum $N(\omega)$. In other terms one is faced with the problem of inverting eq.~(\ref{carson_law}) to obtain the expression of $f(t)$ given $P_f(\omega)$. Then, $n(t)$ can be built iteratively according to eq.~(\ref{series}) with a proper choice of the random parameters $a_k$ and $t_k$.\\

\begin{figure}[htb]
\includegraphics[width=.4\textwidth]{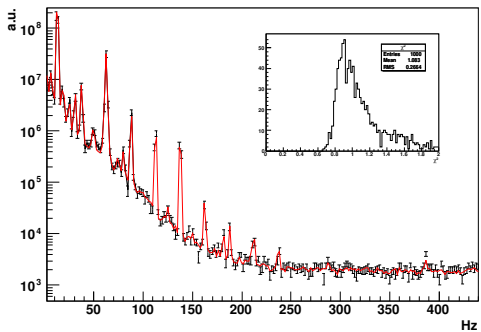}
\caption{Superposition between a simulated (black) and measured (red) bolometric noise power spectrum obtained with a bolometric detector of O(mg) mass~\cite{ale03}. The simulated spectrum was obtained averaging 50 finite time series. Error bars are the average standard deviations of the simulated spectra. The distribution of the $\chi^2$/d.o.f. for a set of simulated spectra is shown in the inset.}
\label{menotre_spectra}
\end{figure}

This approach can be quite difficult whenever the desired spectrum is different from a simple power law and analytical approaches to the determination of $f(t)$ are usually not straight-forward. In fact, the desired noise spectrum often shows complex behaviours in the measured frequency range. An example of such a common situation is shown in fig.~\ref{menotre_spectra} (referring to the output of a few hundred microgram bolometric detector) where we have a complex forest of microphonic lines overimposed to a smoothly varying distribution. The anti-aliasing filter cut frequency is also apparent by the quick drop of the power spectrum in the 100-200 Hz region.\\

The method proposed here consists in the determination of an $f(t)$ which exactly matches a given noise power spectrum. This is essentially based on the choice of a particular (out of an infinite number of possibilities) function $f(t)$ which satisfies eq.~(\ref{carson_law}) and can be applied both to the case in which $N(\omega)$ is known analytically or experimentally measured.\\

Let us begin with the discrete case of a sampled sequence $f[k]$. If we consider a finite number L of samples and assume ergodicity, then the power spectrum can be approximated by an ensemble average over a sufficiently large number of finite sequences $f_L[k]$ of fixed length L according to
\begin{equation}
P_f[k] \simeq <|\mathcal{F}(f_{L}[k])|^2>
\label{powerspectrum}
\end{equation}
where $\mathcal{F}$ stands for the Discrete Fourier Transform (DFT) operator.
Since the sequence elements $f_L[k]$ are supposed to be real, the corresponding power spectrum is a real even function and only half of its values are independent. In fact all the informations concerning the phases are lost in the quadrature and this is the reason why the inversion of eq.~(\ref{powerspectrum}) does not admit a unique solution $f_L[k]$.\\ 
The simplest solution to eq.~(\ref{powerspectrum}) is obtained by deleting the ensemble average and inverting the resulting expression. Phases can be then randomly added assuming a flat distribution. The result represents the core of this work and can be summarized as follows:
\begin{equation}
F[k] \equiv \sqrt{N[k]}\cdot e^{i\theta_k}
\label{met01}
\end{equation}
\begin{equation}
f[k]=\mathcal{F}^{-1}(F[k])
\label{met02}
\end{equation}

where $N[k]$ is the power spectrum according to which we want to generate our time series (known a priori) and $\theta_k$ are random numbers uniformely distributed between $0$ and $2\pi$. In order to guarantee the condition of reality of $f[k]$ upon inverse DFT the sYmmetry constraint $F[k]=F^*[-k]$  must be imposed.\\

Noise time series can now be built following eq.~(\ref{series}), simply generating the delays $t_k$ according a Poisson distribution:
\begin{equation}
t_k = t_{k-1} -\ln(1-R)/\lambda
\label{delays}
\end{equation}
where R is a random number with a uniform distribution and $\lambda$, which represents the overlap rate of the signals $f[k]$, is the only free parameter of the method and is usually adjusted to improve the quality of the simulated time series.\\ 

As stated above, the choice of the normalization term $\alpha$ partially fixes the arbitraryness in the choice of the amplitudes density function by setting its variance. This can be better understood by following some of the steps in the derivation of the Carson's theorem. Actually, by substituting eq.~(\ref{series}) into the the power spectrum definition we obtain: 
\begin{eqnarray}
N(\omega) = \left< \left| \mathcal{F} \left[\sum_{k} a_k f(t-t_k)\right] \right|^2 \right>  = \nonumber \\
=\left< \left| F(\omega) \right|^2 \cdot \sum_{ij} a_i a_j e^{i\omega(t_j-t_i)} \right>
\label{car01}
\end{eqnarray}
The last sum can then be separated into two different terms, one for $i=j$, the other for $i\neq j$. It can be shown that the latter is proportional to $F(0)$ and therefore vanishes since we required $f(t)$ to average to zero. We obtain therefore:\\
\begin{equation}
N(\omega) = \left< \left| F(\omega)\right|^2\right>  \left<\sum_{k=1}^{L} |a_k|^2\right>
\label{car02}
\end{equation}
%%%%%%%%%%%%%%%%%%%%%%%%
%%%%%%%%%%%%%%%%%%%%%%%%

\begin{figure}[htb]
\includegraphics[width=.4\textwidth]{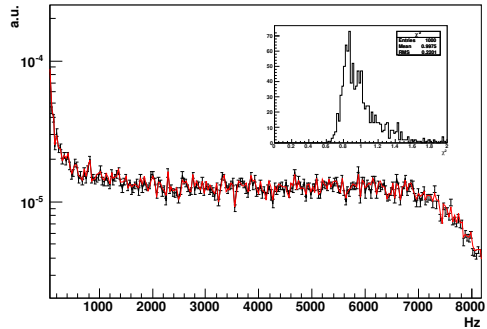}
\caption{Superposition between a simulated (black) and measured (red) jfet noise power spectrum. The simulated spectrum was obtained averaging 50 finite time series. Error bars are the average standard deviations of the simulated spectra. The distribution of the $\chi^2$/d.o.f. for a set of simulated spectra is shown in the inset.}
\label{fet_spectra}
\end{figure}

\begin{figure}[htb]
\includegraphics[width=.4\textwidth]{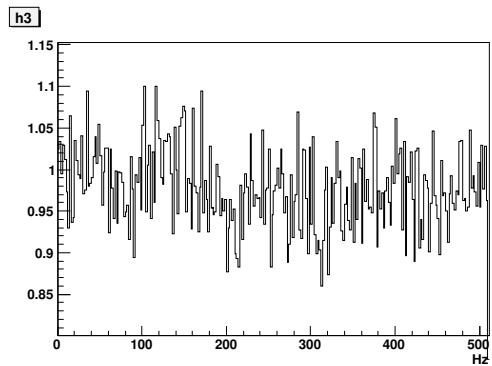}
\caption{Ratio between the simulated and measured bolometric noise power spectra shown in Fig.\ref{menotre_spectra}.} 
\label{bolo_ratio}
\end{figure}

\begin{figure}[htb]
\includegraphics[width=.4\textwidth]{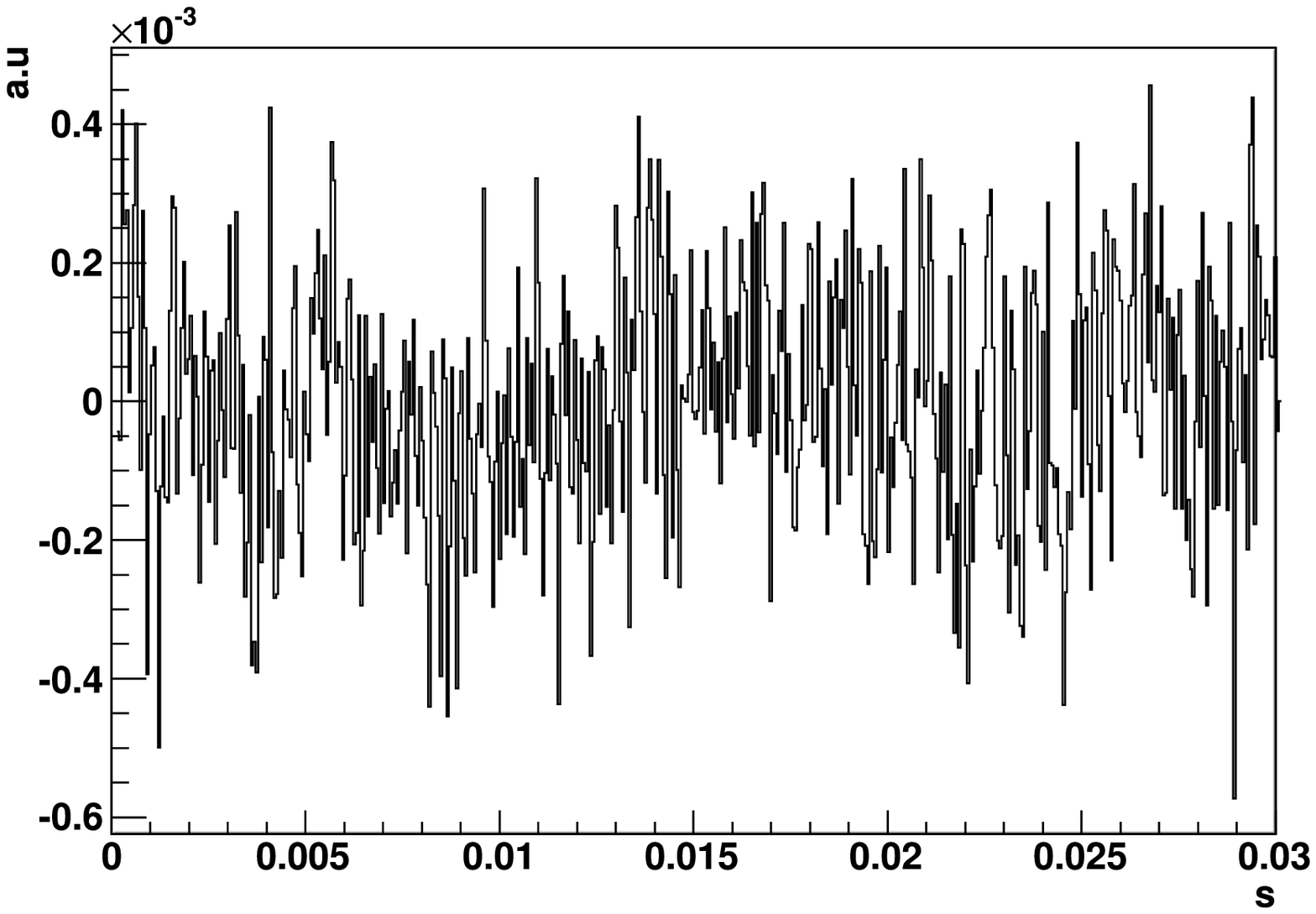}
\includegraphics[width=.4\textwidth]{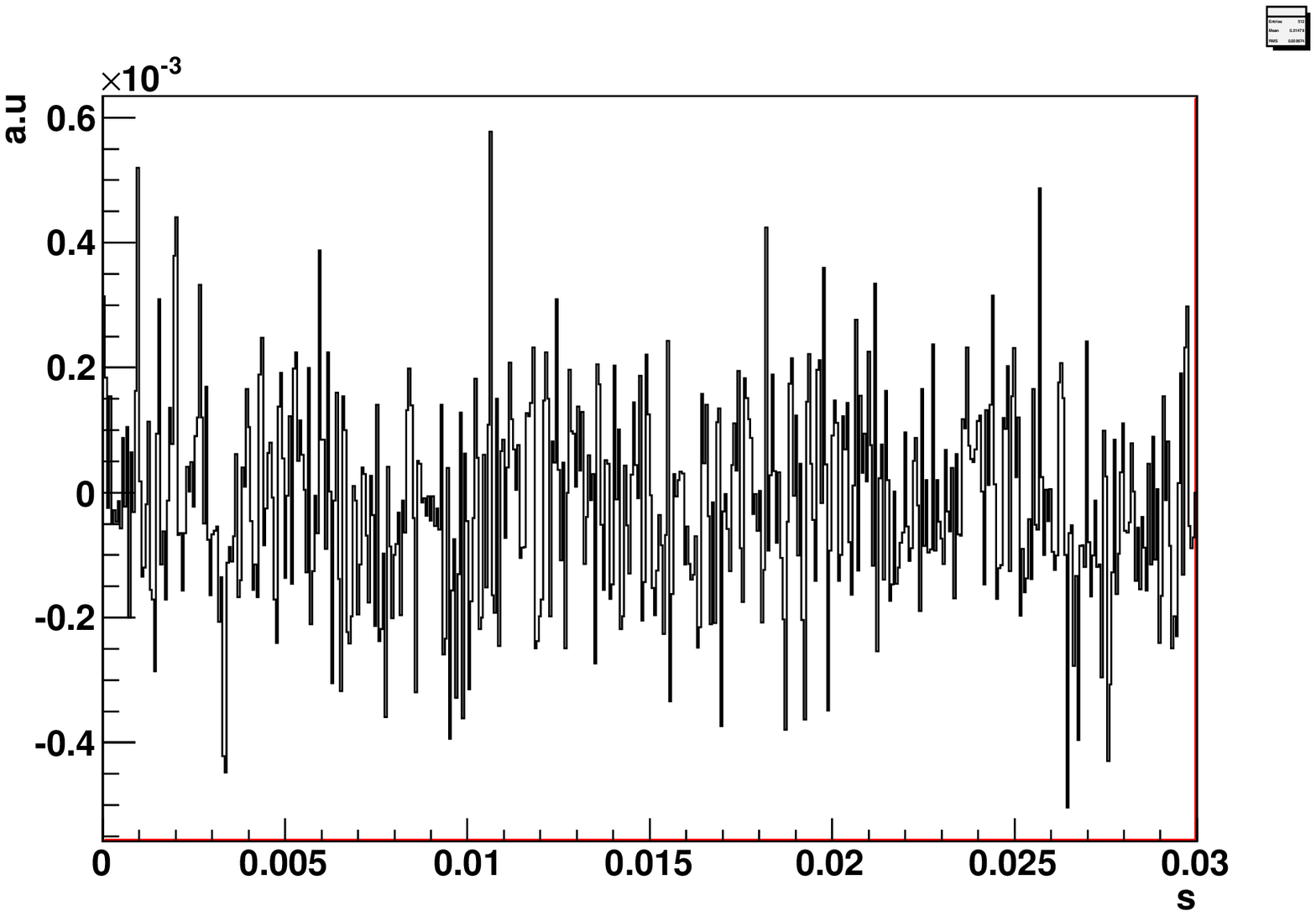}
\caption{Samples of simulated (top) and a real (bottom) noise time series of a jfet}
\label{timedomain}
\end{figure}

\begin{figure}[htb]
\includegraphics[width=.4\textwidth]{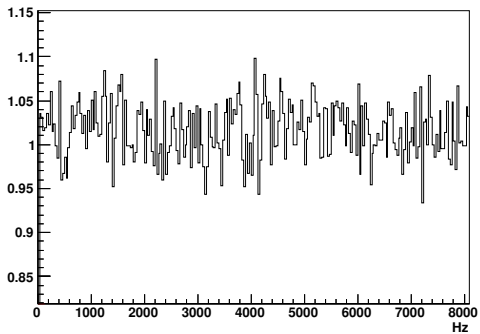}
\caption{Ratio between a simulated and measured jfet noise power spectra shown in Fig.\ref{fet_spectra}.}
\label{fet_ratio}
\end{figure}

\begin{figure}[htb]
\includegraphics[width=.4\textwidth]{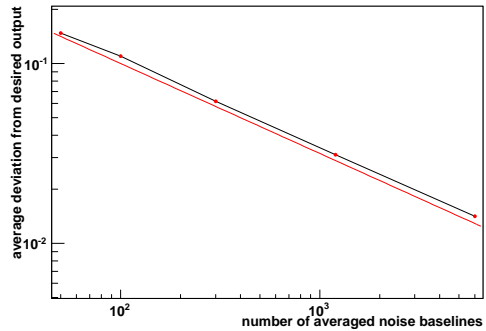}
\caption{Average normalized standard deviation (dots) of the simulated power spectra as a function of the number of simulated time series used for the average process. The black line joins simulated points while the red one is an inverse square root law shown for reference.}
\label{chisquare}
\end{figure}

%%%%%%%%%%%%%%%%%%%%%%%
%%%%%%%%%%%%%%%%%%%%%%%
The averaged sum on the right hand part can be thought as the average of the squared amplitude of each single pulse times their average number $M$ of occurencies. As a result we can express the average power spectrum as: \\
\begin{equation}
N(\omega) = \left< \left| F(\omega)\right|^2\right> M<\left|a\right|^2>
\label{nps_fin}
\end{equation}

By comparing eq.~(\ref{nps_fin}) and eq.~(\ref{carson_law}) we can finally obtain the relation between the average rate $\lambda$, the normalization constant $\alpha$ and the amplitudes variance $|a|^2$ :
\begin{equation}
\left|a\right|^2=\frac {\alpha} {M}=\frac{\alpha}{\lambda T}
\label{rate}
\end{equation}
where $T$ is the finite length of the time series. Any arbitrary distribution with a variance given by eq.~(\ref{rate}) can now be used to generate the random amplitudes $a_k$ and the method is fully defined.\par
We can therefore summarize the most relevant steps for the method implementation as follows:
\\\\\\
\begin{itemize}
\item Given the desired noise power spectrum $N(\omega)$ select the basis function f(t) according to eqs.~(\ref{met01}) and (\ref{met02}). Then eq.~(\ref{carson_law}) fixes the constant $\alpha$ 
\item Generate a set of increasing delays $t_k$ according to a Poisson distribution with average rate $\lambda$. 
\item Generate the random amplitudes $a_k$ according to an arbitrary distribution with variance fixed by eq.(\ref{rate}). 
\item Shift $f(t)$ by $t_k$ imposing a periodicity constraint $f(t)=f(t+T)$ and multiply the result by $a_k$.
\item Sum iteratively over $t_k$ while $\sum t_k<T$
\end{itemize}

The method can be used also when a theoretical (analog) estimate of $F(\omega)$ is known. This is a common situation when the performance of an electronic circuit or of a detector are studied. In this cases the discrete power spectrum F[k] can be in fact obtained by means of a proper frequency warping~\cite{dsp75} of the analog theoretical estimate of the power spectrum.\\

\section{Applications}

We would like to discuss in this section a few examples of application of the method to situations characterized by a complex structure of the desired noise power spectrum. We have chosen to this end two completely different noise sources: a single particle bolometric detector~\cite{fio84,ale99,ale03} and a jfet transistor~\cite{arn04,arn06}. Indeed, being characterized by very different masses, frequency ranges and noise sources, they provide a clear example of the potential and versatility of the method.\par
The concept of a single particle bolometric detector is very simple. It is an ideal calorimeter for which all the energy of the incident radiation is converted to heat giving rise to a temperature rise of the detector's body. Such a temperature variation is then measured by means of a proper transducer which in our case will be represented by a doped thermistor whose resistance varies with temperature. A bias circuit is therefore needed and the observed signals express the voltage change at the thermistor terminals. Working temperatures are as low as few tens of K in order to guarantee measurable temperature variations.
In such working conditions different noise contributions can be recognized: the electronic noise coming from the acquisition chain, the Johnson noise in bias circuit elements, the mechanical vibrations or instabilities of the of the experimental setup which may produce variation in the acquired voltage samples and finally the temperature instabilities due to thermodynamic fluctuations.  The result is a complex  noise power spectrum in which a forest of sharp lines is overimposed to a generally smooth behaviour. Detector details (eg. mass) naturally affect the specific shape of the noise spectra (range, lines, ...) but usually do not alter the general trends described above~\cite{pir06,cuore04}.
Following the common prescriptions~\cite{dsp75} such spectra were obtained by averaging a few hundreds of randomly selected baselines (defined as time windows in the absence of a signal). 
The same averaging procedure was followed to obtain a power spectrum of the output noise of a simple jfet transistor (fig.~\ref{fet_spectra}).  
Following the recipe described in this paper and starting from the power spectra of a small mass bolometer and a jfet, a set of 50 simulated time series was generated. The respective power spectra were obtained according to eq.\ref{powerspectrum} by averaging their DFT's. The result is directly compared in fig.~\ref{menotre_spectra} and \ref{fet_spectra} with the original noise power spectra obtained experimentally. The ratio between the experimental and simulated power spectra is also reported in fig.~\ref{bolo_ratio} and in fig.~\ref{fet_ratio}. Experimental and simulated baselines are also compared in fig.~\ref{timedomain} in the case of the jfet transistor.

In order to allow a more quantitative estimate of the level of agreement between the simulated and the original power spectra, a full set of simulated power spectra was generated in order to evaluate the distribution of the simulated values at each frequency. Each power spectrum was obtained averaging 50 simulated time series. The obtained standard deviations are shown in fig.~\ref{fet_spectra} and fig.~\ref{menotre_spectra}. The distributions of the $\chi^2$/d.o.f. are also shown to demonstrate the correct statistical behaviour.
The fluctuations of the simulated power spectra values depend of course also on the number of time series used for the average process (eq.\ref{powerspectrum}). Such a dependence was explicitely studied by generating a different set of power spectra and evaluating the corresponding average standard deviation from the desired otput, as a function of the number of averaged time series. The result is shown in fig.~\ref{chisquare} and found in excellent agreement with an inverse square root law.


\begin{thebibliography}{99}

\bibitem{carson31}  J.R.~Carson, {\it Bell Syst. Techn. J.} 10 (1931) 374.
\bibitem{dsp75} A.V.~Oppenheim and R.W.Schafer, Digital Signal Processing, Prentice Hall 1975, ISBN 978-0132146357
\bibitem{fio84} E.~Fiorini and T.~Niinikoski, Nucl. Instrum. Meth. 224 (1984) 83.
\bibitem{ale99} A.~Alessandrello et al., Physical Review Letters 82 (1999) 513
\bibitem{ale03} A.~Alessandrello et al., Physical Review Letters 91 (2003) 161802/1
\bibitem{arn04} C.~Arnaboldi et al., Nucl. Instr. and Meth. A 520  (2004) 578
\bibitem{arn06} C.~Arnaboldi et al., Nucl. Instr. and Meth. A 559 (2006) 82
\bibitem{pir06} S.~Pirro, Nucl. Instrum. and Meth. A 559 (2006) 672 sala C
\bibitem{cuore04} C.~Arnaboldi et al., Nucl. Instr. and Meth. A 518  (2004), p. 775

\end{thebibliography}
\end{document}